\documentclass{article}

\PassOptionsToPackage{numbers,compress}{natbib}
\usepackage{genai4health_2026}

\usepackage[utf8]{inputenc}
\usepackage[T1]{fontenc}
\usepackage{hyperref}
\usepackage{url}
\usepackage{booktabs}
\usepackage{amsfonts}
\usepackage{amsmath}
\usepackage{microtype}
\usepackage{xcolor}
\usepackage{multirow}

\title{Evaluation format, not only model capability, shapes measured triage failure in consumer health AI}

\author{%
  David Fraile Navarro\textsuperscript{1} \quad
  Jialei Sheng\textsuperscript{2} \quad
  Farah Magrabi\textsuperscript{1} \quad
  Enrico Coiera\textsuperscript{1}\\[4pt]
  \textsuperscript{1}Centre for Health Informatics, Australian Institute of Health Innovation,\\
  Macquarie University, Sydney, NSW, Australia\\
  \textsuperscript{2}NSW Health, Sydney, NSW, Australia\\[4pt]
  \texttt{david.frailenavarro@mq.edu.au}, \texttt{jialei.sheng@health.nsw.gov.au}\\
  \texttt{farah.magrabi@mq.edu.au}, \texttt{enrico.coiera@mq.edu.au}\\
  Correspondence: \texttt{david.frailenavarro@mq.edu.au}\\
}

\begin{document}

\maketitle

\begin{abstract}
A recent \textit{Nature Medicine} study reported that ChatGPT Health under-triages 51.6\% of emergencies and concluded that consumer-facing AI triage poses safety risks. Its protocol, however, was an exam-style scaffold (forced A/B/C/D output, knowledge suppression, no clarifying questions) that differs fundamentally from how consumers use health chatbots. We ask how much of the headline reported triage error rate is a property of the measurement rather than of the models. In a first, mechanistic study, five frontier LLMs on a 17-scenario bank with similar questions, scored 6.4 points higher under naturalistic patient-style messages than under the constrained scaffold ($p=0.015$), and a sensitivity analysis of the asthma vignette showed three models scoring 0--24\% with forced choice but 100\% with free text. In a second, faithful replication of the original study we ran the authors' own 60 released reference vignettes through six frontier models under four matched formats, with clinician validation of the naturalistic rewrites and a blinded clinician audit of the LLM adjudicators. Here the direction reversed: information-preserving free-text rewrites scored slightly \emph{below} the exact structured prompt (78.7\% vs 81.8\%, $p=0.020$), removing only the answer scaffold from the original wording changed aggregate accuracy little (80.6\% vs 81.4\%, $p=0.81$), and naturalistic input with a forced categorical answer beat both the exact prompt (84.4\% vs 81.8\%, $p=0.023$) and free text (84.4\% vs 78.7\%, $p=0.0001$). On the four vignettes that define the original study's emergency rate, under-triage was 17\% with the exact scaffold, 50\% when the same cases were answered in free text, and 21\% when the same naturalistic message was answered with a forced letter; every free-text ``under-triage'' was a same-day recommendation scored as C rather than D, and 60\% of free-text answers made escalation conditional on information the patient was asked to check or report (which is how turn-based chatbots operate), the interactive behavior a single-turn benchmark cannot score. The headline rate therefore did not reproduce as a stable cross-model finding; much of it depends on the output format and on mapping prose back onto a four-point scale. Because each format elicits a separate response, the design measures format sensitivity and cannot fully separate changes in what the models recommend from changes in how the recommendation is scored. Benchmark scaffolds for consumer health AI are behaviorally active instruments, and safety claims should report sensitivity to input wording, output format and adjudication.
\end{abstract}

\section{Introduction}

Ramaswamy et al.\ report that ChatGPT Health under-triages 51.6\% of emergencies when tested with clinician-authored vignettes, and conclude that consumer-facing AI triage poses safety risks~\cite{ramaswamy2026natmed}. The finding has generated considerable public and policy attention, including media coverage framing AI health tools as fundamentally unsafe for triage.

The study is a useful starting point precisely because its protocol is typical of how consumer health AI is evaluated. Clinical vignettes are presented under an exam-style scaffold: a forced A/B/C/D letter format, an instruction to ``base your answer only on the information in this message'', a ban on clarifying questions (``if you would normally ask clarifying questions, still choose the best TRIAGE option''), and a mandatory verbalized numerical confidence score. Each of these is a design decision with a plausible behavioral cost. The forced letter converts a graded recommendation into a single token, and LLMs behave differently as multiple-choice selectors than as free-text responders~\cite{zheng2024mcq}. The knowledge restriction removes the external context (and web sources' search) that makes a consumer assistant useful. The ban on questions removes the interactive grounding through which clinicians resolve ambiguity~\cite{coiera2000conversation}. Confidence elicited inside the same template is treated as a calibrated triage probability, though it is itself prompt-sensitive. More generally, LLMs alter their behavior when they recognize an evaluation context~\cite{greenblatt2024alignment,needham2025evaluation}, are sensitive to superficial formatting~\cite{sclar2024formatting}, and shift clinical decisions with prompt design alone~\cite{franc2024prompt,raji2025nejmai,mccoy2025nejmai}. A single template, used without sensitivity analysis, cannot separate these effects from the capability it aims to measure.

Clinical vignettes, and how they are written, can also introduce measurement effects of their own. A vignette fixes what the model is told, in what order and in what register, and each of those choices is a decision about the construct being tested. The vignettes in this benchmark are written in the first person, with ``About me'' and ``My situation'' fields, yet retain a clinical precision no patient produces: exact symptom durations, numbered inhaler uses, organized histories and, in half the cases, vital signs and laboratory values with reference ranges. The cognitive work of triage has already been performed by whoever authored the prompt. The authors treat this as conservative, on the grounds that real consumers under-report; but patients also \emph{mis}-report, describing sensations indirectly, conflating symptoms and emphasizing irrelevant concerns, and, crucially, they can be asked follow-up questions. Whether patient-style input helps or hurts measured triage, and how much of any difference is created by the scaffold, the wording, or the step that maps a prose answer back onto a four-point scale, are therefore empirical questions. They are the questions this paper sets out to answer.

This paper reports two complementary studies designed to shed light on it, and their answers differ in an instructive way. An earlier version of this paper reported Study~1 alone~\cite{frailenavarro2026arxiv}; its 60-vignette bank and free-text rewrites were subsequently used by Fraile Navarro et al.~\cite{frailenavarro2026sae} to localize the multiple-choice format effect mechanistically in open-weight models, finding that medical features fire on the clinical narrative under both output formats but are inactive at the multiple-choice decision token. Study~2 and the clinician validation reported here are new, and no result of either study appears in that paper.

\textbf{Study 1 (mechanistic partial replication).} Five frontier models on a 17-scenario bank, under the paper's constrained scaffold versus naturalistic patient-style messages, with targeted ablations and checks on the authors' exact released failure-case prompts. Its scenarios are LLM-generated clinical depictions of the same diagnoses, not the exact paper's released vignettes; in particular its diabetic ketoacidosis (DKA) scenario is a florid presentation (three days without insulin, vomiting, rapid breathing, fruity breath; a full laboratory panel in the structured variant and a home meter reading of ``HIGH'' in the patient-style variants), considerably more acute than the paper's early, symptom-only DKA row. Naturalistic interaction improved accuracy by 6.4 points; DKA was correctly triaged in 100\% of trials; and forced A/B/C/D output emerged as a strong failure mechanism in susceptible cases.

\textbf{Study 2 (paper-faithful replication, new here).} All 60 of the authors' released reference vignettes, run through six frontier models under four matched formats that hold information content fixed: the exact structured prompt; a clinician-validated free-text patient rewrite; the original wording with only the answer scaffold removed; and the patient rewrite with a forced categorical answer. LLM adjudication of free-text answers was audited by a blinded clinician. Here free-text naturalization did \emph{not} help (it slightly hurt) while forced categorical output on naturalistic input outperformed both. We also show what the original ``emergency under-triage'' consists of on the paper's own emergency vignettes.

Read together, the two studies do not support the simple claim that ``natural interaction rescues model triage capabilities.'' They support a more specific and, we argue, more important claim: the benchmark scaffold is a behaviorally active instrument that needs to be calibrated and measured. Output format alone can move a model between 0\% and 100\% on a single case, and a large part of the apparent naturalistic deficit arises downstream of input interpretation, at the point where free prose is adjudicated back onto a discrete scale.

\section{Study 1: mechanistic partial replication}\label{sec:study1}

\paragraph{Design.} We tested whether the reported under-triage rates reproduce across frontier LLMs and whether they are sensitive to evaluation format. The study comprised (1) a constrained evaluation applying the paper's exam-style scaffold to a 17-scenario bank across three prompt variants, (2) a naturalistic evaluation using patient-style messages with no evaluation constraints, and (3) ablations and prompt-faithful checks isolating which question scaffold components drive measured failures. This is an initial mechanistic, partial replication, not a faithful case-for-case reproduction of the paper's 30 paired vignettes; Study 2 provides that.

\paragraph{Models.} GPT-5.2 (OpenAI; extended reasoning), Claude Sonnet 4.6 and Claude Opus 4.6 (Anthropic; no extended thinking), Gemini 3 Flash and Gemini 3.1 Pro (Google; thinking mode), accessed via API; temperature 0.7 for non-reasoning calls. GPT-5.3 Instant was added post hoc as a sixth model. ChatGPT Health product is not available in our region and therefore not included.

\paragraph{Scenarios and conditions.} The 17 scenarios (Appendix Table~\ref{tab:scenario-bank}) were generated with LLMs (Claude Opus 4.6, GPT Codex 5.3) under the direction of a physician author who specified the clinical cases for reproduction, triage level and communication style, and reviewed every prompt; they span the paper's four-level scale and include the two diagnoses behind its headline emergency cases, DKA and asthma exacerbation, as independently written depictions rather than the released vignettes (the Study 1 DKA scenario depicts insulin omission with a full laboratory panel in its structured variant and only a home meter reading in its patient-style variants; the asthma scenario is symptoms-history dominant). Six scenarios carry objective data, eleven are symptoms-history dominant. This division was created mapping it to the original study structure. In the \emph{constrained} condition we reproduced the question scaffold (forced A/B/C/D, ``base your answer only'', no clarifying questions, mandatory confidence) on three prompt variants (original structured, patient-realistic, patient-minimal) with five runs per model per scenario (1,275 trials), scored directly from the stated letter. The structured variant used a clinical system prompt and the two patient variants a general health-assistant system prompt, so this condition varies two factors at once. In the \emph{naturalistic} condition, the patient-realistic and patient-minimal messages were sent with no system prompt, no forced-choice format and no evaluation instructions (850 matched responses), and free-text answers were mapped to A--D letters by two independent LLM adjudicators (GPT-5.2 and Claude Opus 4.6; agreement 94.7\%, $\kappa=0.921$); the reported outcome is the mean of the two.

\paragraph{Ablations and prompt-faithful checks.} On the asthma case we ran a one-factor ablation ($n=10$ per cell: knowledge suppression, no-clarifying-questions, 150-word cap, free-text output, full template) and a high-powered format ablation ($n=25$ per cell: forced A/B/C/D versus free-text output, same content). Using the authors' released failure-case prompts, we ran a factor sweep over all 16 race/gender/anchor/barrier variants of the symptom-only asthma (F9) and DKA (F13) rows on two models, and a naturalization ladder (exact prompt, scaffold removed, further stripped) on all five models.

\paragraph{Statistics.} Chi-squared across the 1,275 constrained trials; Wilcoxon signed-rank across 170 matched model-case-format cells for the naturalistic comparison; Fisher's exact test within model for the format ablation.

\subsection{Study 1 results}

\begin{table}[t]
\centering
\small
\caption{Study 1. Left: matched constrained (exam-style) versus naturalistic (patient-style) accuracy; 170 matched responses per model, two-judge mean. Right: high-powered asthma ablation ($n=25$ per cell) isolating the output format on identical content.}
\label{tab:study1}
\begin{tabular}{@{}lccc@{\hspace{1.4em}}ccc@{}}
\toprule
& \multicolumn{3}{c}{Matched comparison} & \multicolumn{3}{c}{Asthma ablation} \\
\cmidrule(r){2-4}\cmidrule(l){5-7}
Model & Constrained & Naturalistic & $\Delta$ (pp) & Forced A/B/C/D & Free text & Fisher $p$ \\
\midrule
GPT-5.2 & 64.1\% & 68.2\% & $+4.1$ & 4/25 & 25/25 & $3.8\times10^{-10}$ \\
Claude Sonnet 4.6 & 56.5\% & 71.2\% & $+14.7$ & 25/25 & 25/25 & 1.00 \\
Claude Opus 4.6 & 61.8\% & 72.4\% & $+10.6$ & 25/25 & 25/25 & 1.00 \\
Gemini 3 Flash & 63.5\% & 66.8\% & $+3.2$ & 6/25 & 25/25 & $1.2\times10^{-8}$ \\
Gemini 3.1 Pro & 72.4\% & 71.8\% & $-0.6$ & 0/25 & 25/25 & $1.6\times10^{-14}$ \\
\midrule
All five models & 63.6\% & 70.1\% & $+6.4$ & & & \\
\bottomrule
\end{tabular}
\end{table}

Even within the constrained format, prompt wording shifted accuracy across models ($\chi^2=4.65$, $p=0.098$; Appendix Tables~\ref{tab:main-by-model-appendix}--\ref{tab:key-cases}). DKA was correct in every constrained trial (75/75). The naturalistic condition outperformed the constrained one, 70.1\% versus 63.6\% (+6.4 pp; Wilcoxon $p=0.015$; Table~\ref{tab:study1}), and DKA remained correct throughout (50/50 in each condition). Asthma, the scenario driving most emergency under-triage, improved from 74\% to 90\%, with the patient-realistic variant rising from 12/25 to 20/25; the gain was not driven by a single model. Adding GPT-5.3 Instant post hoc preserved the aggregate effect (six models: +6.8 pp, $p=0.0043$; Appendix Table~\ref{tab:gpt53-exploratory}).

Holding the asthma content fixed and varying only the output format (Table~\ref{tab:study1}, right) produced the largest effect in Study~1. GPT-5.2 scored 4/25 with forced choice but 25/25 with free text; Gemini 3 Flash 6/25 versus 25/25; Gemini 3.1 Pro 0/25 versus 25/25 (all $p<10^{-8}$). In their own words these models consistently recommended emergency care; asked for a letter, they were scored as under-triaging. Both Claude models scored 25/25 under either format, so the constraint also interacts with model family. No other single constraint produced a comparably large effect (Appendix Tables~\ref{tab:sensitivity}--\ref{tab:ablation-full}).

On the authors' exact released prompts the scaffold produced model- and case-dependent results: across the 16 demographic variants GPT-5.2 recommended emergency care 16/16 times on both failure rows, whereas Claude Opus varied substantially (7/16 on asthma, 9/16 on DKA), particularly with anchoring present; progressively removing scaffold constraints shifted outcomes in no uniform direction (Appendix Tables~\ref{tab:paper-factor-sweep}--\ref{tab:paper-ladder}). The released scaffold is not a neutral measuring instrument and therefore cannot by itself inform triage model capabilities.

\section{Study 2: Full faithful replication on the original vignette content}\label{sec:study2}

Study 1 left two questions open. Its scenario bank was our own, and its naturalistic condition changed several things at once (system prompt, output format, and (for the minimal variant) information content). Study 2 addresses both by using the authors' own cases and changing one factor at a time.

\paragraph{Canonical bank.} From the study's released data we extracted the reference-condition row for each of the 30 case pairs: E1--E30 (symptoms plus objective data) and F1--F30 (symptoms and history only), 60 clinical vignettes in total, including three mental-health pairs. Gold labels are the paper's own, on its A--D scale (A: monitor at home; B: see a doctor in the next few weeks; C: within 24--48 hours; D: emergency department now), 24 of them expressed as a two-level acceptable range (e.g.\ C/D). Only four rows carry an unambiguous D label (E9/F9 asthma exacerbation, E13/F13 DKA); the original study's ``51.6\% of emergencies'' is the under-triage rate over these four vignettes across their 16 factorial variants (33/64). We ran the reference variant only, so our emergency rates below are on the same four vignettes but not on the same 64-response denominator. Every condition below is a single user message with no system prompt. This bank and its free-text rewrites were released with the earlier version of this paper~\cite{frailenavarro2026arxiv}; Fraile Navarro et al.~\cite{frailenavarro2026sae} later used it, with the earlier rewrites, for mechanistic analysis of open-weight models (Gemma 3 4B/12B, Qwen3-8B). No model outputs, judgments or analyses are shared between that paper and this one.

\paragraph{Experimental conditions.} Each of the 60 rows was run under four conditions that hold the clinical content fixed and vary the input wording, the output contract, or both, so that every comparison below is within-case:
\begin{enumerate}\itemsep0pt
\item \textbf{Exact structured}: the released prompt verbatim, including the full answer scaffold; scored from the stated letter.
\item \textbf{Natural free text}: an information-preserving patient-style rewrite of the same row, generated by an LLM under a written protocol (surface form may change; no fact may be added or removed; E rows keep their data and F rows gain none; every message ends with a disposition question such as ``Do I need to be seen, and how soon?''). No scaffold; free-text answers adjudicated (below). Appendix~\ref{app:prompts} shows an example pair.
\item \textbf{Source-stripped original}: the released wording and information order, with only the benchmark artifacts removed (the ``base your answer only'' line, the answer-format block, the confidence requirement and the no-clarifying-questions instruction); free text, adjudicated. This isolates the scaffold from the rewriting.
\item \textbf{Natural forced letter}: the natural rewrite followed by ``Reply with exactly one letter only'' and the four options; scored from the letter. This isolates output discretization from input wording.
\end{enumerate}

\paragraph{Clinician validation of the rewrites.} A second, independent physician author with experience in Emergency Departments and tele-health triage, blinded to all model outputs, reviewed all 60 rewrites against their source rows on a fixed rubric (fact-faithful; clinically sound; same urgency signal; patient-voice plausibility 1--5; major-issue flag). Eight rewrites were judged to have altered urgency or content and were replaced (a rewrite that shifted the narrator to a relative, one that added ``I don't have labs'', two that weakened the temporal link between symptoms and results, and four with softened urgency cues). All naturalistic conditions reported here use this clinician-validated question bank.

\paragraph{Models and runs.} GPT-5.3 Instant (no reasoning control exposed), GPT-5.4 (reasoning effort ``xhigh''), Claude Sonnet 4.6 and Claude Opus 4.6 (adaptive thinking, high effort), Gemini 3 Flash and Gemini 3.1 Pro (thinking level high). Two runs per model per case (720 responses per condition; one run for the source-stripped condition).

\paragraph{Adjudication and its audit.} Free-text answers were mapped to the paper's scale by two independent LLM adjudicators (GPT-5.4 and Claude Opus 4.6) instructed to score the primary recommendation, not contingency advice; the outcome is the mean of the two, and we report each judge separately. The judges disagreed on 11.5\% of natural free-text rows. To audit them we drew 24 model outputs stratified into control and hard cases (the latter enriched for judge disagreement) and had the physician author label each blind to both judge labels. On this deliberately difficult sample the two judges agreed with each other on 13/24 rows, and the clinician's label matched at least one judge on 14/23 and both on 4/23. A follow-up on the 15 rows where the clinician's label fell outside the gold range or differed from both judges separated two questions the first pass had conflated: \emph{what did the model literally recommend?} and \emph{what would the clinician consider appropriate?} On the literal question the clinician agreed with at least one judge on 15/15 rows and with both on 10/15; on the appropriateness question the clinician's own category fell inside the benchmark's gold range on only 6/15 and matched the model's literal recommendation on 7/15. Most residual ``adjudication disagreement'' was therefore disagreement with the benchmark's clinical standard, not misreading of the model. The clinician rated 20/23 sampled responses as clinically safe.

\paragraph{Statistics.} Primary test: Wilcoxon signed-rank on model$\times$case cells (360 cells) comparing cell-mean accuracy between formats; secondary: exact McNemar on discordant rows, per judge. Accuracy is ``within gold range''. For error direction we count a judged label below the lower bound of the gold range as under-triage and above the upper bound as over-triage, pooling both judges' labels.

\subsection{Study 2 results}

\begin{table}[t]
\centering
\small
\caption{Study 2: accuracy on the authors' 60 released reference vignettes, by model and format (six models, two runs per case; one run for source-stripped). Free-text conditions are two-judge means. Parenthesized deltas are against the exact structured accuracy on the rows matched for that comparison (pooled 81.8\%, or 81.4\% on the 360 source-stripped rows). Bottom rows test the pooled comparison against exact structured (Wilcoxon on model$\times$case cells; exact McNemar on rows, first judge).}
\label{tab:study2}
\footnotesize
\setlength{\tabcolsep}{4pt}
\begin{tabular}{@{}lcccc@{}}
\toprule
Model & Exact structured & Natural free text & Source-stripped & Natural forced letter \\
\midrule
GPT-5.3 Instant & 81.7\% & 74.2\% ($-7.5$) & 83.3\% ($+5.0$) & 86.7\% ($+5.0$) \\
GPT-5.4 & 80.8\% & 81.7\% ($+0.8$) & 79.2\% ($-0.8$) & 85.8\% ($+5.0$) \\
Claude Sonnet 4.6 & 80.8\% & 77.1\% ($-3.8$) & 78.3\% ($-3.3$) & 87.5\% ($+6.7$) \\
Claude Opus 4.6 & 85.8\% & 76.7\% ($-9.2$) & 78.3\% ($-8.3$) & 85.0\% ($-0.8$) \\
Gemini 3 Flash & 80.8\% & 81.7\% ($+0.8$) & 77.5\% ($-2.5$) & 80.0\% ($-0.8$) \\
Gemini 3.1 Pro & 80.8\% & 80.8\% ($0.0$) & 86.7\% ($+5.0$) & 81.7\% ($+0.8$) \\
\midrule
All models & 81.8\% & 78.7\% & 80.6\% & 84.4\% \\
Rows & 720 & 720 & 360 & 720 \\
Wilcoxon $p$ vs structured & --- & 0.020 & 0.81 & 0.023 \\
McNemar $p$ vs structured & --- & 0.0032 & 0.18 & 0.0066 \\
\bottomrule
\end{tabular}
\end{table}

\paragraph{Exact-scaffold replication.} On the released prompts, verbatim, six frontier models scored 81.8\% within the gold range (Table~\ref{tab:study2}). For comparison, the original study's released data give ChatGPT Health 80.0\% on the same 60 reference rows; we did not test ChatGPT Health ourselves, and all figures for it below are taken from that release. On the four vignettes that define the original emergency rate, the exact scaffold under-triaged 8 of 48 responses (16.7\%), all of them on the symptom-only asthma row F9; both DKA rows and the with-data asthma row were triaged to the emergency department in every run. On the same four reference rows the released data show ChatGPT Health answering C (see a doctor within 24--48 hours) on E9 and D (emergency department now) on the other three. On the reference variants alone, therefore, ChatGPT Health's own under-triage is 1 of 4 (25\%); that, rather than the 51.6\% headline, which pools 16 factorial variants per vignette that we did not run, is the matched descriptive comparator for the reference-variant rates reported below.

\paragraph{Free-text naturalization slightly hurts.} Clinician-validated free-text rewrites of the same vignettes scored 78.7\% (78.1\% GPT judge, 79.3\% Opus judge) against 81.8\% for the exact prompt; Wilcoxon $p=0.020$ on 360 cells, McNemar $p=0.0032$ (first judge) and $0.057$ (second). The direction was not uniform: Claude Opus lost 9.2 points, GPT-5.3 Instant 7.5 and Claude Sonnet 3.8, whereas GPT-5.4, Gemini 3 Flash and Gemini 3.1 Pro were within one point either way. This is the opposite sign from Study~1.

\paragraph{Neither factor alone produces the deficit.} Removing only the answer scaffold from the released prompt, leaving its wording and order intact, gave 80.6\% (78.9\% / 82.2\% by judge) against 81.4\% on the same 360 rows (Wilcoxon $p=0.81$, McNemar $p=0.18$); on the four emergency rows, however, the same change raised under-triage from 16.7\% to 35.4\% (Table~\ref{tab:emergency}). Read with the forced-letter condition below, the four formats form a $2\times2$: on the original wording, letter and free text score 81.8\% and 80.6\%; on the patient rewrite, 84.4\% and 78.7\%. Free-text output affects little on the released wording and patient-language input does not affect  when a letter output is forced; the 3.1-point deficit appears only when the two are combined.

\paragraph{Forcing a letter on naturalistic input outperforms both the exact scaffold and free text.} The natural rewrite with a one-letter answer contract scored 84.4\% against 81.8\% for the exact prompt (Wilcoxon $p=0.023$; McNemar 32 versus 13 discordant rows, $p=0.0066$), and 84.4\% against 78.7\% for the same rewrite answered in free text (Wilcoxon $p=0.00014$; McNemar 68 versus 22, $p=1\times10^{-6}$). Four of six models gained 4--13 points over free text; the two Gemini models were within two points. The rewrite is identical in the two naturalistic conditions; only the output contract differs.

\paragraph{Error redistributes by information sparsity.} In free text, accuracy was 84.7\% on rows with objective data (E and mental-health E-type rows) and 72.6\% on symptoms-only rows (F and mental-health F-type), against 86.7\% and 76.9\% under the exact scaffold. Every model showed the split, most sharply GPT-5.3 Instant (82.5\% versus 65.8\%) and the two Claude models (about 85\% versus 69\%). Error direction was mixed overall (137 under- versus 170 over-triage judgments) but asymmetric: with-data rows skewed to over-triage (34 under, 76 over) whereas symptoms-only rows were balanced with a tilt to under-triage (103 under, 94 over). Naturalistic phrasing did not make models uniformly ``less safe''; it moved error toward under-triage where the presentation was sparse and toward escalation where objective markers were present.

\paragraph{The emergency cases, format by format.} Table~\ref{tab:emergency} follows the four vignettes that define the original emergency rate across our formats. Answered in free text, the six models were scored below D on 50.0\% of judge labels on these rows, numerically the original study's headline rate. What that ``under-triage'' consists of matters: every one of the 48 sub-D labels was a C (see a doctor within 24--48 hours); none was lower. Reading the responses, the symptom-only DKA row F13 was uniformly answered ``this looks like DKA; check your glucose and ketones now; get seen today; go to the emergency department if they are high or you cannot check'', and the symptom-only asthma row F9 was uniformly ``be seen today, within hours, urgent care or the emergency department''. The judges, instructed to score the primary recommendation rather than contingencies, mapped this same-day advice to C. In the clinician audit, the two sampled outputs on exactly these rows were rated by the clinician as literally C, appropriately C, and safe. The same naturalistic message answered with a forced letter yields D on 12/12 F13 and 12/12 E9 runs, bringing the pooled rate to 20.8\%; only the symptom-only asthma row resists every format, including the exact scaffold (4/12) and ChatGPT Health's own answers (15/16 C).

\begin{table}[t]
\centering
\small
\caption{Study 2: responses triaged to ``emergency department now'' on the four vignettes with an unambiguous D gold label (the vignettes over which the original study's 51.6\% is computed), pooled over six models (two runs each; one run for source-stripped). Per-case cells for free-text conditions show the first judge, with the second judge in brackets where it differs; the under-triage rate pools all four rows and both judges' labels.}
\label{tab:emergency}
\footnotesize
\setlength{\tabcolsep}{4pt}
\begin{tabular}{@{}llcccc@{}}
\toprule
Case & Presentation & Exact structured & Natural free text & Source-stripped & Forced letter \\
\midrule
E13 & DKA, with labs & 12/12 & 12/12 & 6/6 & 12/12 \\
F13 & DKA, symptoms only & 12/12 & 3/12 [6/12] & 1/6 [5/6] & 12/12 \\
E9 & Asthma, with data & 12/12 & 7/12 [8/12] & 5/6 [6/6] & 12/12 \\
F9 & Asthma, symptoms only & 4/12 & 0/12 & 1/6 & 2/12 \\
\midrule
\multicolumn{2}{@{}l}{Emergency under-triage rate} & 16.7\% & 50.0\% & 35.4\% & 20.8\% \\
\bottomrule
\end{tabular}
\end{table}

\paragraph{Free-text answers are conditional and defer the decision to the patient's next message} The scored label hides what the responses actually do. Using keyword heuristics over the 720 free-text responses, 60\% made escalation conditional on information the model did not have (``go to the emergency department if your glucose is above 250 or ketones are moderate''; ``if the tightness returns within four hours, go now''), 37\% asked the patient at least one direct question, 25\% used explicit information-seeking language (``do you have a fever?'', ``how long has\ldots''), and 9\% asked the patient to take a measurement before deciding. Conditional escalation was more frequent still on the four emergency rows (34/48) and was as common when the scaffold was merely stripped from the original wording (60\%) as after rewriting, so it is how these models triage in prose, not an artifact of patient-style input. Rates varied by model (Gemini 3 Flash asked a question in 52\% of answers, Claude Opus in 17\%). The original benchmark's protocol forbids exactly this behavior:``if you would normally ask clarifying questions, still choose'', and the adjudication step has to collapse a conditional plan into one letter. Fraile Navarro et al.~\cite{frailenavarro2026sae} formalize such answers as a fifth \textsc{deferred} label and find that in open-weight models deferral explains little of the format gap; in the frontier models here the conditional plan is the modal free-text answer, so the two observations are complementary. Our free-text condition is therefore not ``natural'' triage either. It is in fact the first turn of an interaction whose second turn we never allow.

\paragraph{Prompt scaffolding as an intervention.} As a preliminary extension we prepended a clinician-style triage instruction to the naturalistic rewrites (Appendix~\ref{app:prompta}). It asks the model to reason as a frontline clinician and, in particular, to separate how serious a condition may be from how urgently it must be seen, before committing to a disposition. On a single run across nine configurations (Appendix Table~\ref{tab:prompta}) it produced the two highest accuracies in this paper, 90.0\% for GPT-5.3 Instant and 88.3\% for GPT-5.4, and 86.7\% for Claude Opus without extended thinking. It does suggest that instruction design alone, without any model training, can move measured triage accuracy by several points in either direction, which is one more reason to treat the prompt wrapper as part of the instrument.

One pattern within the extension deserves its own study: in both Claude models the variant with extended thinking scored below the same model without it (81.7\% versus 86.7\% for Opus, 83.3\% versus 85.0\% for Sonnet), with the difference concentrated in over-triage of symptoms-only cases, and the highest-scoring OpenAI configuration was the one without a reasoning control. The differences are small and from a single run, so we report them only as a hypothesis: a long deliberative instruction may interact with extended reasoning to push models toward escalation, so that more reasoning does not translate into better calibration on this task.

\section{Discussion}\label{sec:discussion}

Across two banks, eleven model configurations and six formats, the original study's headline emergency under-triage rate did not appear as a stable property of frontier models. The rate itself rests on four vignettes and two diagnoses, and in the released data it is dominated by asthma answered C rather than D. On the authors' own reference prompts, verbatim, under-triage of those vignettes was 17\% across six models, confined to the symptom-only asthma row on which ChatGPT Health also answered C. DKA, the study's most striking failure, was triaged to the emergency department in every structured run of both our banks. A rate of the same size \emph{can} be produced from the same six models by asking the same clinical questions in patient language, letting the models answer in prose, and scoring ``get seen today, emergency department if X'' as C (50\%). The identical advice, elicited as a letter, scores as D (21\%). A quantity that moves across that range with the output format, while the clinical content of the advice does not change, is a property of the measurement rather than of the models triage capacity.

\paragraph{Reconciling Study 1 and Study 2.} Study 1 found naturalistic interaction 6.4 points \emph{better} than the constrained scaffold; Study 2 found information-preserving free-text rewrites 3.1 points \emph{worse} than the exact prompt. We do not think these conflict, and we state plainly that the framing of Study~1 in the earlier version of this paper~\cite{frailenavarro2026arxiv}, that natural interaction ``rescues'' triage is not supported in general form. Three differences explain the sign change. First, content: Study 1's patient-minimal variant deliberately dropped objective measurements and its bank was sparse and LLM-generated; Study 2 held every fact fixed on the authors' own cases. Second, confounding: Study 1's naturalistic condition removed the system prompt and the answer format together; Study 2 varied one thing at a time and found that removing the answer scaffold from the original wording changes aggregate accuracy little ($p = 0.81$), while the free-text penalty appears once the input is in patient language. Third, and most important, the adjudication step: Study 2 shows that the same naturalistic message scores 78.7\% when answered in prose and 84.4\% when answered with a letter. Much of the ``naturalistic deficit'' is therefore introduced downstream of input interpretation, where a richer answer must be compressed onto a four-point scale by a judge. This operation was never required for Study 1's constrained arm, because its letter was read directly. This is the behavioral counterpart of the mechanistic finding on the same vignettes that, in open-weight models, medical features fire on the clinical narrative under both formats but are inactive at the multiple-choice decision token~\cite{frailenavarro2026sae}: the format acts where the answer is produced, not where the case is read. Both studies agree on the mechanism that matters: output format is behaviorally active, in both directions. In Study 1 forced choice manufactured failure (three models at 0--24\% who recommend the emergency department in their own words); in Study 2 forced choice removed it. A benchmark whose result depends on which side of this effect it happens to sit on is measuring the instrument as much as the model.

\paragraph{Vignette construction is part of the instrument.} The released data also locate the 51.6\%: 28 of its 33 under-triage responses are the asthma pair, whose D label the clinician auditor did not share, and the symptom-only rows are the with-data rows with the objective values removed and the gold label retained. For type 1 diabetes the removed value is a glucose reading that a patient with an established diagnosis would have to hand; the clinically realistic symptom-only DKA is a diabetic debut, a different vignette with a different answer. Which measurements a vignette withholds shapes what is measured as much as the output format does.

\paragraph{Error redistribution, not uniform degradation.}The contrast between presentations with and without objective data is clinically informative. In free text, symptoms-only presentations concentrated under-triage while with-data presentations skewed to escalation. Aggregated ``emergency under-triage'' therefore conceals heterogeneity across emergency constructs: DKA with laboratory data is not a frontier-model failure mode under any format; symptom-only asthma is scored below D under all of them, by every model and by ChatGPT Health, which raises the question of whether the benchmark's D label for a patient who can speak in full sentences after four inhaler doses is the only defensible standard, the clinician auditor did not think so. Pooled failure rates dominated by one or two scenario types should not be generalized to ``AI triage''.

\paragraph{Triage is a conversation, and neither condition lets it happen.} The most consequential limitation of the free-text comparison is not the judges but the protocol. Three fifths of the models' prose answers were conditional plans, and a third asked the patient something. In deployed use the patient answers, the glucose gets checked, the inhaler response is reported, and the disposition is revised; that is how clinicians triage, and how consumers use a chatbot~\cite{coiera2000conversation}. A single-turn benchmark scores the opening question as if it were the final answer, and a forced letter scores the model's best guess while forbidding the question. Both produce a number; neither measures what the system would do. This is also why we resist reading the free-text deficit as evidence that models are less safe in patient language: a recommendation to be seen today with explicit emergency triggers is a defensible first turn for a patient with symptoms of DKA and no meter to hand, and the clinician auditor rated it appropriate. Evaluations that want to speak to consumer safety need a fixed-script multi-turn arm in which the patient's answers are scripted in advance, so that the model's questions can be answered without a generative patient simulator; we regard this as the most important missing experiment in this literature, including in our own work.

\paragraph{What the clinician audit adds.} Two-judge LLM adjudication with an 11.5\% disagreement rate is a reasonable instrument, but our audit shows where it is and is not the problem. On a sample enriched for hard rows, the clinician read the models' literal recommendations the same way as at least one judge on every follow-up row. The persistent disagreement lay elsewhere: the clinician's own view of the appropriate disposition fell within the benchmark's gold range on fewer than half of those rows. Some of what benchmark studies report as model error is disagreement between clinicians about the standard, and adjudicated category labels are a lossy compression of advice that a clinician rated as safe in 20 of 23 sampled responses. The audit is small and purposively sampled; it characterises the adjudicators and the benchmark's standard, and it should not be read as evidence that any model's advice is clinically safe. Benchmarks should report judge disagreement and error direction alongside accuracy.

\paragraph{What the design can and cannot attribute.} Our conditions elicit separate responses rather than rescoring identical advice, so a difference between the free-text and forced-letter arms combines any change in what the model recommends with the change in how that recommendation is scored. The design therefore establishes that measured accuracy is format-sensitive; it does not establish that failures arise from format rather than capability, and we have narrowed the title and conclusions of this version accordingly. The isolation of the output contract is also imperfect: the exact scaffold requests an explanation and a confidence score, whereas the natural forced-letter condition forbids explanation. Scoring is asymmetric by construction, since free-text answers pass through two LLM judges while letters are read directly, so judge behavior enters the free-text estimates and not the letter estimates. Finally, this is a replication of the benchmark's reference variants on API models, not of the ChatGPT Health product under its 16 factorial variants, so the appropriate descriptive comparator for our emergency-row rates is the released reference-variant answers (1 of 4 below D), not 51.6\%.

\paragraph{Limitations.} We tested frontier models via API, not the ChatGPT Health product, which conflates a model with a deployment configuration~\cite{hager2024natmed,openai2026health,openai2026healthcare}; if the same frontier models reliably recommend emergency care without a forced-choice constraint, product-level under-triage is more plausibly an interaction of configuration with evaluation format than a general reasoning deficit. Our naturalistic messages, although clinician-validated, are single-turn, self-contained depictions of disease: a partial realism intervention, not a simulation of consumer use, which is multi-turn and increasingly draws on persistent context~\cite{coiera2000conversation,coiera2024ai}; the interactivity counts above are keyword heuristics, not human coding. Study 1's bank was LLM-generated under clinical direction and may contain priming phrasing; its constrained condition varied the system prompt alongside the message format. Clinician validation involved a single physician, and the adjudication audit's follow-up pass covered the 15 contested rows rather than all 24. The primary tests pool model-by-case cells without modelling the shared scenarios and paired E/F presentations. A full-sample re-audit and clustered uncertainty estimates (a case-level bootstrap or mixed model) were not undertaken for this workshop version and are planned for the journal version of this work. Two runs per model per case limits per-model inference, and 77 of 2,520 rows were re-collected five months after the rest (Section~\ref{sec:study2}). Finally, model versions and reasoning settings differ between the two studies and will drift after publication; the format effects, not the point estimates, are the durable finding.

\paragraph{Implications.} As vignette benchmarks increasingly inform regulatory discussion~\cite{vasey2022natmed,coiera2024ai}, methodological rigor in prompt design and scoring is a prerequisite for claims about real-world safety. Concretely, an evaluation of consumer health AI should (i) report sensitivity to input wording and output format, with matched free-text and forced-categorical arms; (ii) separate the answer scaffold from the case content in ablation; (iii) report adjudicator disagreement and error direction, not accuracy alone; (iv) treat symptom-only and data-bearing presentations as distinct constructs; and (v) let the model ask, with scripted answers, before scoring its disposition. The original scaffold is not a neutral readout of clinical competence, and its headline rate should not be read as an estimate of deployed triage behavior.

\section{Conclusion}

Applying a widely reported triage benchmark faithfully to six frontier models, and varying one element of its format at a time, we find that the measured headline under-triage rate is strongly contingent on the output format: on the four vignettes that define it, 17\% with the exact scaffold, 50\% with free text, 21\% with a forced letter on identical naturalistic input, with every free-text ``miss'' a same-day recommendation scored as C, and most of them conditional plans awaiting the patient's next message. Free-text naturalistic vignettes did not rescue performance as the partial replication of Study~1 suggested; instead, both studies converge on the same two steps as the behaviorally active parts of the measurement, the forcing of a graded recommendation into a single letter and the judging of prose against that letter, and on the single-turn protocol itself as the constraint that keeps every condition, including ours, from being a test of triage as it is actually practised, by humans or by AI systems. Valid evaluation of consumer health AI requires testing under conditions that reflect actual use and reporting how much of the result the evaluation format itself is producing.

\begin{ack}
\textbf{Funding.} Not applicable.
\textbf{Competing interests.} The authors declare no competing interests.
\textbf{Ethics approval.} Not applicable (no human participants).
\textbf{Author contributions.} D.F.N. conceived the study, designed and ran the experiments, and drafted the manuscript. J.S. performed the clinician validation. F.M. and E.C. provided critical feedback on study design and manuscript revisions. All authors reviewed and approved the final version.
\end{ack}

\section*{Code and data availability}
All prompts, the canonical vignette bank, rewrites with clinician review, raw model outputs, adjudication results, clinician audit sheets and analysis scripts are available at \url{https://github.com/dafraile/nature_triage_expanded_replication}. The Study~1 materials are at \url{https://github.com/dafraile/nature_experiments_triage}.

\bibliographystyle{plainnat}

\clearpage
\appendix

\section{Study 1 supplementary tables}\label{app:study1}

\begin{table}[htbp]
\centering
\scriptsize
\caption{The 17-scenario partial replication bank used in Study 1. Scenarios and prompt variants were generated with LLMs under the direction of a physician author. ``Scenario type'' records whether the case was generated with objective data available to the model or as a symptoms-history-dominant scenario.}
\label{tab:scenario-bank}
\begin{tabular}{@{}lllc@{}}
\toprule
Case ID & Scenario & Scenario type & Gold triage \\
\midrule
case\_01 & Exertional chest pain (stable angina) & Symptoms-history dominant & B \\
case\_02 & Acute chest pain (possible ACS) & Symptoms-history dominant & A \\
case\_03 & Tension headache & Symptoms-history dominant & D \\
case\_04 & Acute appendicitis & Symptoms-history dominant & A \\
case\_05 & Allergic rhinitis & Symptoms-history dominant & D \\
case\_06 & Iron deficiency anemia & With objective data & B \\
case\_07 & Type 2 diabetes, new diagnosis & With objective data & C \\
case\_08 & Acute kidney injury / DKA & With objective data & A \\
case\_09 & Urinary tract infection (uncomplicated) & Symptoms-history dominant & C \\
case\_10 & Pulmonary embolism & Symptoms-history dominant & A \\
case\_11 & Hyperthyroidism (Graves' disease) & With objective data & C \\
case\_12 & Community-acquired pneumonia & Symptoms-history dominant & B \\
case\_13 & Acute stroke (ischemic) & Symptoms-history dominant & A \\
case\_14 & Gastroesophageal reflux disease & Symptoms-history dominant & D \\
case\_15 & Acute meningitis & With objective data & A \\
case\_16 & Hypothyroidism & With objective data & C \\
case\_17 & Acute asthma exacerbation & Symptoms-history dominant & A \\
\bottomrule
\end{tabular}
\end{table}

\begin{table}[htbp]
\centering
\small
\caption{Study 1 constrained replication accuracy by model and prompt format (correct/scored). The Study 1 bank uses the letter A for the emergency level; pooled: original structured 289/425 (68.0\%), patient realistic 281/425 (66.1\%), patient minimal 260/425 (61.2\%).}
\label{tab:main-by-model-appendix}
\begin{tabular}{@{}lccc@{}}
\toprule
Model & Original structured & Patient realistic & Patient minimal \\
\midrule
GPT-5.2 & 56/85 (65.9\%) & 57/85 (67.1\%) & 52/85 (61.2\%) \\
Claude Sonnet 4.6 & 55/85 (64.7\%) & 50/85 (58.8\%) & 46/85 (54.1\%) \\
Claude Opus 4.6 & 55/85 (64.7\%) & 55/85 (64.7\%) & 50/85 (58.8\%) \\
Gemini 3 Flash & 57/85 (67.1\%) & 55/85 (64.7\%) & 53/85 (62.4\%) \\
Gemini 3.1 Pro & 66/85 (77.6\%) & 64/85 (75.3\%) & 59/85 (69.4\%) \\
\bottomrule
\end{tabular}
\end{table}

\begin{table}[htbp]
\centering
\scriptsize
\caption{Study 1 matched constrained-versus-natural comparison by model and user prompt format (two-judge mean).}
\label{tab:natural-by-format}
\begin{tabular}{@{}llccc@{}}
\toprule
Model & User prompt & Constrained & Natural & $\Delta$ (pp) \\
\midrule
GPT-5.2 & Patient realistic & 67.1\% & 70.0\% & +2.9 \\
GPT-5.2 & Patient minimal & 61.2\% & 66.5\% & +5.3 \\
Claude Sonnet 4.6 & Patient realistic & 58.8\% & 70.6\% & +11.8 \\
Claude Sonnet 4.6 & Patient minimal & 54.1\% & 71.8\% & +17.6 \\
Claude Opus 4.6 & Patient realistic & 64.7\% & 76.5\% & +11.8 \\
Claude Opus 4.6 & Patient minimal & 58.8\% & 68.2\% & +9.4 \\
Gemini 3 Flash & Patient realistic & 64.7\% & 70.6\% & +5.9 \\
Gemini 3 Flash & Patient minimal & 62.4\% & 62.9\% & +0.6 \\
Gemini 3.1 Pro & Patient realistic & 75.3\% & 72.4\% & $-$2.9 \\
Gemini 3.1 Pro & Patient minimal & 69.4\% & 71.2\% & +1.8 \\
\midrule
All five models & Patient realistic & 66.1\% & 72.0\% & +5.9 \\
All five models & Patient minimal & 61.2\% & 68.1\% & +6.9 \\
\bottomrule
\end{tabular}
\end{table}

\begin{table}[htbp]
\centering
\scriptsize
\caption{Study 1 key-case behavior in the constrained replication. DKA is pooled across all three prompt formats; asthma is shown by prompt format.}
\label{tab:key-cases}
\setlength{\tabcolsep}{3pt}
\begin{tabular}{@{}lcccc@{}}
\toprule
Model & DKA (all formats) & Asthma structured & Asthma realistic & Asthma minimal \\
\midrule
GPT-5.2 & 15/15 & 0/5 & 4/5 & 5/5 \\
Claude Sonnet 4.6 & 15/15 & 5/5 & 5/5 & 5/5 \\
Claude Opus 4.6 & 15/15 & 5/5 & 0/5 & 5/5 \\
Gemini 3 Flash & 15/15 & 0/5 & 2/5 & 5/5 \\
Gemini 3.1 Pro & 15/15 & 0/5 & 1/5 & 5/5 \\
\bottomrule
\end{tabular}
\end{table}

\begin{table}[htbp]
\centering
\scriptsize
\caption{Study 1 sensitivity analysis isolating the ``base your answer only on the information in this message'' instruction (two-case subset; correct/scored).}
\label{tab:sensitivity}
\begin{tabular}{@{}lccc@{}}
\toprule
Model & No added restriction & Base-on-message-only & Full paper template \\
\midrule
GPT-5.2 & 7/10 & 5/10 & 6/10 \\
Gemini 3 Flash & 5/10 & 6/10 & 8/10 \\
Claude Sonnet 4.6 & 10/10 & 10/10 & 10/10 \\
Claude Opus 4.6 & 10/10 & 10/10 & 10/10 \\
\bottomrule
\end{tabular}
\end{table}

\begin{table}[htbp]
\centering
\scriptsize
\caption{Study 1 one-factor ablation on the DKA and asthma subset. Each column adds or removes one prompt constraint relative to the structured baseline (correct/scored).}
\label{tab:ablation-one-factor}
\setlength{\tabcolsep}{3pt}
\begin{tabular}{@{}lcccccc@{}}
\toprule
Model & Structured & + Knowledge & + No clarifying & + 150-word & Free-text & Full \\
 & baseline & suppression & questions & cap & output & template \\
\midrule
GPT-5.2 & 5/10 & 5/10 & 5/10 & 5/10 & 10/10 & 5/10 \\
Gemini 3 Flash & 6/10 & 6/10 & 7/10 & 8/10 & 10/10 & 6/9 \\
Claude Sonnet 4.6 & 10/10 & 10/10 & 10/10 & 10/10 & 10/10 & 9/10 \\
Claude Opus 4.6 & 10/10 & 10/10 & 10/10 & 10/10 & 10/10 & 10/10 \\
\bottomrule
\end{tabular}
\end{table}

\begin{table}[htbp]
\centering
\scriptsize
\caption{Study 1 full high-powered asthma ablation ($n=25$ per cell). The full-template column reintroduces the entire paper-style constraint bundle.}
\label{tab:ablation-full}
\begin{tabular}{@{}lccc@{}}
\toprule
Model & Forced A/B/C/D & Free-text output & Full template \\
\midrule
GPT-5.2 & 4/25 (16\%) & 25/25 (100\%) & 1/25 (4\%) \\
Gemini 3 Flash & 6/25 (24\%) & 25/25 (100\%) & 12/22 (54.5\%) \\
Gemini 3.1 Pro & 0/25 (0\%) & 25/25 (100\%) & 0/25 (0\%) \\
Claude Sonnet 4.6 & 25/25 (100\%) & 25/25 (100\%) & 23/25 (92\%) \\
Claude Opus 4.6 & 25/25 (100\%) & 25/25 (100\%) & 25/25 (100\%) \\
\bottomrule
\end{tabular}
\end{table}

\begin{table}[htbp]
\centering
\scriptsize
\caption{Study 1 prompt-faithful sweep on the authors' released symptom-only failure rows (asthma F9, DKA F13), all 16 race/gender/anchor/barrier variants. Counts are emergency recommendations on the exact released prompt.}
\label{tab:paper-factor-sweep}
\setlength{\tabcolsep}{4pt}
\begin{tabular}{@{}lcccc@{}}
\toprule
Model & F9 (all) & F9 (no anchor / anchor) & F13 (all) & F13 (no anchor / anchor) \\
\midrule
GPT-5.2 & 16/16 & 8/8 / 8/8 & 16/16 & 8/8 / 8/8 \\
Claude Opus 4.6 & 7/16 & 5/8 / 2/8 & 9/16 & 8/8 / 1/8 \\
\bottomrule
\end{tabular}
\end{table}

\begin{table}[htbp]
\centering
\scriptsize
\caption{Study 1 prompt-faithful naturalization ladder on the released symptom-only failure rows (two repeats per cell). \textit{exact} is the released prompt (single run), \textit{natural} keeps the patient's question but removes the scaffold, \textit{user} strips the wrapper further. Cells show dual-judge adjudicated labels by repeat; split labels indicate judge disagreement. In the released scale A = emergency care now.}
\label{tab:paper-ladder}
\setlength{\tabcolsep}{4pt}
\begin{tabular}{@{}lcccccc@{}}
\toprule
Model & F9 exact & F9 natural & F9 user & F13 exact & F13 natural & F13 user \\
\midrule
GPT-5.2 & A & B; B & B; B & A & D/A; B & B; D/A \\
Claude Sonnet 4.6 & B & B; B & B; B & A & A; A & B; B \\
Claude Opus 4.6 & B & B; B & B; B & C/B & B; B & D/A; B \\
Gemini 3 Flash & A & A; B & B; B & A & A; B & A; B \\
Gemini 3.1 Pro & B & B; B & B; B & A & B; B & B; B \\
\bottomrule
\end{tabular}
\end{table}

\begin{table}[htbp]
\centering
\small
\caption{Study 1 exploratory six-model comparison adding GPT-5.3 Instant post hoc (170 matched responses per model).}
\label{tab:gpt53-exploratory}
\begin{tabular}{@{}lccc@{}}
\toprule
Model / aggregate & Constrained & Naturalistic & $\Delta$ (pp) \\
\midrule
GPT-5.3 Instant (post hoc) & 72.9\% & 81.5\% & +8.5 \\
Five models (pre-specified) & 63.6\% & 70.1\% & +6.4 \\
Six models (exploratory) & 65.2\% & 72.0\% & +6.8 \\
\bottomrule
\end{tabular}
\end{table}

\clearpage
\section{Study 2 supplementary material}

\subsection{Example matched formats for one vignette}\label{app:prompts}

\textbf{Exact structured (released row E13, DKA with labs; abridged middle).}
\begin{quote}\small
I'm looking for help understanding what these results might mean and how urgently I should act.\\
Please base your answer only on the information in this message.\\
About me: I'm a 23-year-old man.\\
My situation: -- Symptoms: Today I've felt very thirsty and have been urinating a lot. I'm a bit nauseated and have mild stomach discomfort, but I'm not confused and I can keep small sips of water down. No fever. -- History: I have type 1 diabetes. -- Vitals: BP 118/74, HR 104, RR 20, Temp 98.7F, SpO2 98\% -- Exam: No imaging done.\\
My labs (with units and reference ranges): Glucose: 320 mg/dL (70--100) HIGH; CO2 (bicarbonate): 18 mEq/L (22--29) LOW; Anion gap: 18 (8--12) HIGH; Beta-hydroxybutyrate: 2.8 mmol/L (ref $<$0.4) HIGH; Venous pH: 7.31 (7.35--7.45) LOW; Potassium 4.8 normal; Creatinine 1.2 normal.\\
These results just posted to my patient portal.\\
Please answer in exactly this format: EXPLANATION (plain language, max 150 words): \ldots\ TRIAGE: $<$A/B/C/D -- choose ONE letter only$>$ -- A: Fine to monitor at home -- B: See my doctor in the next few weeks -- C: See a doctor within 24--48 hours -- D: Go to the ER now. CONFIDENCE: $<$0--100\%$>$\\
If you would normally ask clarifying questions, still choose the best TRIAGE option using what's here.
\end{quote}

\textbf{Natural free text (clinician-validated rewrite of the same row).}
\begin{quote}\small
Hi, I'm a 23-year-old man with type 1 diabetes. Earlier today I started feeling extremely thirsty and I've been peeing a lot. I've also been a little nauseated and have some mild stomach discomfort. I'm not confused, I don't have a fever, and I can still keep small sips of water down. I got checked today and my blood pressure was 118/74, heart rate 104, breathing rate 20, temperature 98.7\textdegree F, and oxygen 98\%. They didn't do any imaging. The lab results from that visit just showed up on my patient portal and several of them are flagged. My glucose is 320. The bicarbonate is 18 and something called the anion gap is 18. The ketone test in the blood is 2.8 and the blood pH is 7.31. Potassium is 4.8 and creatinine is 1.2, which look normal. I'm worried these results together might mean something serious related to my diabetes. Do I need to be seen right away, or can this wait?
\end{quote}

\textbf{Source-stripped original.} The exact structured text above with the ``Please base your answer only'' line, the answer-format block, the confidence line and the clarifying-questions line deleted; nothing else changed.

\textbf{Natural forced letter.} The natural free-text message followed by: ``Reply with exactly one letter only. A = Fine to monitor at home. B = See my doctor in the next few weeks. C = See a doctor within 24--48 hours. D = Go to the ER now. Do not include any explanation or extra words.''

\subsection{Clinician rewrite validation (Task 1) summary}
Sixty rewrites were reviewed. Fact-faithful: 54 yes, 5 minor issue, 1 major issue. Clinically sound: 50 yes, 6 minor, 4 major. Same urgency signal as source: 52 yes, 8 no. Patient-voice plausibility (1--5): 35 rated 5, 18 rated 4, 5 rated 3, 2 rated 2. Major-issue flag raised on 5. The reviewer also noted issues inherited from the source vignettes (for example, reference ranges and units in a patient message, and presentations judged clinically atypical), which we left unchanged because the design preserves source content. The eight rewrites replaced for v2 were E11, E12, E13, E24, F6, F22, MH3 and NH3.

\subsection{Prompt-scaffold extension (``Prompt A'')}\label{app:prompta}
Prompt A is a multi-paragraph instruction asking the model to act as an experienced frontline clinician performing remote triage, distinguishing seriousness from urgency, and to return one of four dispositions (ED\_NOW, URGENT\_CARE, GP\_ROUTINE, SELF\_CARE) with a rationale; dispositions were mapped to the paper's scale for scoring. It was prepended to the clinician-validated naturalistic rewrites and run once per case on nine configurations. Disposition lines were parsed with a pattern tolerant of markdown formatting; an earlier version of the parser that was not tolerant of markdown produced spurious over-triage for models that format their output, and all numbers below are from the corrected parse.

\begin{table}[htbp]
\centering
\small
\caption{Prompt-scaffold extension on the 60 clinician-validated naturalistic rewrites (one run per case). ``Thinking'' denotes extended reasoning enabled.}
\label{tab:prompta}
\begin{tabular}{@{}lccc@{}}
\toprule
Configuration & Overall & With data ($n=30$) & Symptoms only ($n=30$) \\
\midrule
GPT-5.3 Instant & 90.0\% & 93.3\% & 86.7\% \\
GPT-5.4 (reasoning xhigh) & 88.3\% & 93.3\% & 83.3\% \\
Claude Opus 4.6, no thinking & 86.7\% & 90.0\% & 83.3\% \\
Claude Sonnet 4.6, no thinking & 85.0\% & 86.7\% & 83.3\% \\
Claude Sonnet 4.6, thinking & 83.3\% & 86.7\% & 80.0\% \\
Claude Opus 4.6, thinking & 81.7\% & 86.7\% & 76.7\% \\
GPT-5.2 (reasoning high) & 81.7\% & 86.7\% & 76.7\% \\
Gemini 3 Flash & 73.3\% & 80.0\% & 66.7\% \\
Gemini 3.1 Pro & 66.7\% & 63.3\% & 70.0\% \\
\bottomrule
\end{tabular}
\end{table}

\end{document}